\documentclass[aps,preprint]{revtex4}%
\usepackage{amsfonts}
\usepackage{amssymb}
\usepackage{graphicx}
\usepackage{color}
\usepackage{amsmath}%
\providecommand{\U}[1]{\protect\rule{.1in}{.1in}}

\begin{document}
\preprint{ }
\title[ ]{The mechanism that drives electrostatic solitary waves to propagate in the Earth's magnetosphere and solar wind}
\author{M. S. Afify$^{1,a,\ast}$, R. E. Tolba$^{2,b}$, W. M. Moslem$^{2,3,c}$}
\affiliation{$^{1}$Department of Physics, Faculty of Science, Benha University, Benha, Egypt}
\affiliation{$^{2}$Centre for Theoretical Physics, The British University in Egypt (BUE),
El-Shorouk City, Cairo, Egypt}
\affiliation{$^{3}$Department of Physics, Faculty of Science, Port Said University, Port
Said, Egypt}

\begin{abstract}
The origin of the solitary waves in the Earth's magnetosphere and the solar wind, particularly close to the magnetic reconnection regions, is still an open question. For that purpose, we used the fluid model to obtain the mechanism behind the formation of solitary waves in space plasmas with a variety of components. Our findings reveal that at the separatrix of the magnetic reconnection in the Earth's magnetotail, the counter-ejected electron beam leads to a broadening of the pulse amplitude. While increasing the electron beam positive velocity causes a bit increase in the soliton amplitude. Moreover, we observed that the soliton profile is very sensitive to the plasma parameters like density and temperature. The calculated electric field using the current model and the measured electric field in the Earth's magnetosphere and solar wind is highlighted. \\
\textbf{keywords:} Electrostatic solitary waves; Earth's magnetosphere; Solar wind

----------------------------------------------------------

$^{\text{*}}$Corresponding author: M. S. Afify

$^{\text{a}}$mahmoud.afify@fsc.bu.edu.eg

$^{\text{b}}$tolba\_math@yahoo.com

$^{\text{c}}$wmmoslem@hotmail.com
\end{abstract}
\startpage{1}
\endpage{24}
\date{\today}
\maketitle

\section{Introduction}

Plasma represents most of the matter in the universe, which includes the nebulae, stellar interiors and atmospheres, and interstellar space \cite{Thorne}. 
Since the plasma is an ionized gas, the motion of the charged particles is governed by the electric and magnetic fields. 
Plasma groups with opposite line directions of magnetic field give rise to the generation of magnetic reconnection when these lines are met \cite{2,3}. 
This phenomena can be observed in coronal mass ejections (CMEs) from the Sun, solar flares, active galactic nuclei jets, x-ray flares in pulsar wind nebulae, Earth's magnetosphere, and laboratory experiments such as laser-matter interaction and nuclear fusion \cite{4,5}.
Indeed, magnetic reconnection is a characteristic plasma process where the magnetic energy is converted to thermal and kinetic energy. 
This mechanism is responsible for heating the ambient plasma and forming jets \cite{re}.

Moreover, magnetic reconnection has negative effects on space weather, satellites, pipelines and power grid, and fusion energy according to the following reasons: (i) The direct interaction between the solar CMEs, i.e. a billion tons or so of plasma from the sun, and the Earth's magnetosphere lead to an event of space weather called geomagnetic storms or substorm. 
These storms are responsible for a strong disturbance in the Earth's magnetosphere. 
The magnetic reconnection is driven during the geomagnetic storms that cause heating of the background plasma. 
The aurora lights at the Earth's poles that result when the accelerated particles penetrate down to Earth's atmosphere could produce also large magnetic disturbances \cite{re2}. (ii) the occurrence of magnetic reconnection near the sunspots is the mechanism behind the propagation of billions joules of energy that called solar flares. 
When more flares reached at Earth's atmosphere, it enhances the activity of the geomagnetic storms. 
Moreover, it is absorbed by the Earth's ionosphere generating more charged particles. 
Therefore, the density of the plasma increases where electrical currents can drive through conductors causing damage to the power grid and pipelines on the ground. 
Also, increasing the density of the plasma in the ionosphere leads to increasing the resistance in the ionosphere, which is the mechanism behind the veer off satellites from the desired course that impacts the communications and GPS signals. 
Solar flares can corrupt solar cells on satellites which
is harmful to astronauts \cite{rh}. 
(iii) In fusion devices such as tokamak, the magnetic reconnection which is formed between the twisted magnetic field weaken the confinement process required to produce the energy \cite{r1}. 
On the other hand, magnetic reconnection can provide the human with a huge amount of energy, for instance, the large flare can supply sufficient energy for the entire world for several years. 
The conversion of the magnetic energy which is a result from the generation of the magnetic reconnection into jets can be utilized in plasma thrusters \cite{r2}. 
Thus, the magnetic reconnection is an important frontier area of the science driving space weather.

The connection between different types of waves that have been observed in reconnecting plasmas and the production mechanisms of these waves remains unclear so far. 
In the magnetic reconnection experiment, the lower-hybrid drift instability had been detected \cite{r3}.
The cross-field currents are responsible for the propagation of such high-frequency drift wave instability in the presence of a gradient of the density and magnetic field. 
The existence of this instability leads to producing anomalous resistivity and heating the background plasma \cite{r4}.
Further, the occurrence of the lower hybrid waves in the Earth's magnetotail during the magnetic reconnection had been reported (more details are in Ref. \cite{r5}). 
The magnetic reconnection between the hot magnetospheric plasma and the cold dense magnetosheath plasma is the mechanism behind the perturbation of the separatrix regions. 
This instability changes the electron distribution giving rise to the propagation of whistler waves toward the X-line \cite{r6}. 
Wilder et al. \cite{r7} showed that the generation of whistler waves at the separatrix regions during the magnetic reconnection could be assigned to the Landau resonance with a beam of electrons moving along the magnetic field or the instability of whistler mode anisotropy. The latter has a significant role in regulating the heat flux in the solar wind \cite{Micera2021}. 
Moreover, several waves such as Langmuir waves, upper hybrid waves, and electron cyclotron waves had been observed in the separatrix of magnetotail reconnection and magnetosheath separatrices \cite{r8, r9, r10}.

Electrostatic solitary waves have been ubiquitous in space plasmas since 1982. (\cite{r11,Afify2021} and references therein) . 
Pickett et al. \cite{r12} reported the propagation of bipolar and tripolar solitary waves from the bow shock to the magnetopause. 
They argued that the two-stream instability is the mechanism behind the generation of soliton waves. 
In 1998, Ergun et al. \cite{r13} observed the propagation of fast soliton waves in the downward current region. 
They concluded that these waves are responsible for the existence of large-scale parallel electric fields and diverging electrostatic shock waves resulting from the acceleration of electrons.
Moreover, the solitary waves had observed in the solar wind and more details about the mechanisms that contributed to the turbulence, single soliton, and soliton train can be found in Ref. \cite{r14}. 
Interestingly, the generation of soliton waves in the magnetotail and the magnetopause is associated with the production of magnetic reconnection that caused a disturbance of the electron distribution (see e.g., Refs. \cite{r11, r15, r16, r17, r18, r19} and references therein).

To understand the mechanisms those are responsible for the formation of soliton waves in regions of magnetic reconnections, many efforts have been done \cite{r20, r21, r22, r23, r24}. 
Omura et al. \cite{r20} used computer simulation to investigate the growth of soliton waves at the magnetic reconnection.
They found that four possible mechanisms are responsible for the generation of soliton waves which are bump-on-tail instability, bistream instability, warm bistream instability, and weak-beam instability. 
In 2010, the particle-in-cell simulation and analytic model was employed by Che et al. \cite{r21}. 
They supposed that the Buneman and lower hybrid instabilities lead to the existence of soliton waves during the magnetic reconnection. 
Moreover, several works highlighted the role of Buneman in driving the solitary waves associated with the magnetic reconnection, and further details can be found in Refs. \cite{r22, r23, r24, r25, r26}. 
In the separatrix region, Divin et al. \cite{r27} showed that the extension of solitary waves in the perpendicular direction of the magnetic field should be controlled via the interaction between electron Kelvin-Helmholtz instability and Buneman instability. 
The propagation of large amplitude soliton waves in the reconnection jet region of the Earth's magnetotail was investigated by Lakhina et al. \cite{r28} utilizing the Sagdeev pseudopotential technique. 
This study predicted the generation of four fast and slow soliton waves, where the potential profile of these pulses has been examined. 
Rufai et al. \cite{r29} discussed the observation of small amplitude electron acoustic soliton waves in the electron diffusion region of the Earth's magnetopause. 
Their results revealed that the soliton waves propagated at supersonic speeds and both the temperature and the density of the background hot electrons had a significant effect on the profile of soliton waves. 
Recently, Kamaletdinov et al. \cite{r30} suggested a mechanism behind the existence of slow soliton waves in the reconnection current sheets. 
Unfortunately, their kinetic analysis is unable to determine the origin of the slow electron-hole waves since the electrons are colder than ions. 
The technique of Sagdeev potential had been employed by Lakhina et al. \cite{Lakhina2018, Lakhina2021} to discuss the observation of the electrostatic waves in the lunar wake and the solar wind.

In the present work, we examine the dynamics of electrostatic solitary waves in the Earth's magnetosphere and the solar wind. 
This paper is arranged according to the following: the formulation of the problem is presented in Sec. II, while the possible solutions of the evolution equation are discussed in Sec. III. 
The impact of the plasma parameters on the soliton pulse and the application of our model in the Earth's magnetosphere and the solar wind are presented by Secs. IV and V, respectively. Finally, the summary of this work and the future perspectives are reported in Sec. VI.

\section{Physical Model}

Consider a system consisting of positive ions, an electron beam, and superthermal electrons with a Kappa distribution, which is unmagnetized, collisionless, warm, and adiabatic. The following are the basic equations that describe the propagation of ion-acoustic waves in dimensionless variables:

\begin{equation}
\frac{\partial n_{i}}{\partial t}+\frac{\partial}{\partial x}(n_{i}u_{i})=0,
\tag{1}%
\end{equation}%
\begin{equation}
\frac{\partial u_{i}}{\partial t}+u_{i}\frac{\partial u_{i}}{\partial x}%
+\frac{\partial\phi}{\partial x}+3\sigma_{i}n_{i}\frac{\partial
n_{i}}{\partial x}=0, \tag{2}%
\end{equation}
for positive ions, 

\begin{equation}
\frac{\partial n_{e}}{\partial t}+\frac{\partial}{\partial x}(n_{e}u_{e})=0,
\tag{3}%
\end{equation}%
\begin{equation}
\frac{\partial u_{e}}{\partial t}+u_{e}\frac{\partial u_{e}}{\partial x}%
-\mu_{e}\frac{\partial\phi}{\partial x}+3\sigma_{e}\mu_{e}n_{e}\frac{\partial n_{e}%
}{\partial x}=0, \tag{4}%
\end{equation}
for electron beam, while the background electrons are described by the kappa distribution as
\begin{equation}
n_{h}=\left(  1-\frac{\phi}{k_{h}-3/2}\right)  ^{-(k_{h}-1/2)}. \tag{5}%
\end{equation}
The positive ions, electrons and electron beam are coupled through Poisson's equation
\begin{equation}
\frac{\partial^{2}\phi}{\partial x^{2}}=\delta_{e}n_{e}+n_{h}-\delta_{i}n_{i}.
\tag{6}%
\end{equation}

Here, $\mu_{e}=m_{i}/m_{e}$, $\delta_{i}=n_{i0}/n_{h0}$, and $\delta_{e}=n_{e0}/n_{h0}$, where $n_{i0}$ is the unperturbed number density of ions, $n_{e0}$ is the unperturbed number density of electron beam and $n_{h0}$ is the unperturbed number density of superthermal electrons. 
$\sigma_{i}=T_{i}/T_{h}$ and $\sigma_{e}=T_{e}/T_{h}$ where $T_{i}$, $T_{h}$, and $T_{e}$ are the ions, background electrons, and electron beam temperatures, respectively. 
$u_{i}$ and $u_{e}$ are the velocities of ions and electron beam which are normalized by the ion-acoustic speed $C_{s}=\left(  m_{i}/k_{B}T_{h}\right)^{-1/2}$, $k_{B}$ is the Boltzmann constant, and $m_{i}$ is the proton mass. 
The electrostatic potential $\phi$ is normalized by the thermal potential $\left(k_{B}T_{h}/e\right)$, $e$ is the electronic charge. 
The time is normalized by the inverse of ion plasma frequency $\omega_{p}^{-1}=\left(  n_{h0}e^{2}/\varepsilon_{0}m_{i}\right) ^{1/2}$, and the space coordinate $x$ is normalized by the Debye length
$\lambda_{D}=\left(  k_{B}T_{h}/\varepsilon_{0}n_{h0}e^{2}\right)  ^{1/2}$.

To investigate the one-dimensional ion-acoustic waves, we used the standard reductive perturbation method \cite{6rt} to reduce the system of fluid Eqs. (1)--(6) to one nonlinear evolution equation. 
The independent variables can be stretched as
\begin{equation}
\xi=\varepsilon^{1/2}\left(  x-V_{p}t\right)  \text{ and }\tau=\varepsilon
^{3/2}t, \tag{7}%
\end{equation}
where $\varepsilon$ is a small real parameter (i.e $0<\varepsilon\ll1$) and $V_{p}$ represents the wave phase velocity. 
The dependent variables are expanded as
\begin{align}
n_{i}  &  =1+\varepsilon n_{i1}+\varepsilon^{2}n_{i2}+\varepsilon^{3}%
n_{i3}+...,\text{ }u_{i}=\varepsilon u_{i1}+\varepsilon^{2}u_{i2}%
+\varepsilon^{3}u_{i3}+...,\text{ }\nonumber\\
\text{ }\phi &  =\varepsilon\phi_{1}+\varepsilon^{2}\phi_{2}+\varepsilon
^{3}\phi_{3}+...,\text{ }n_{e}=1+\varepsilon n_{e1}+\varepsilon^{2}%
n_{e2}+\varepsilon^{3}n_{e3}+....,\tag{8}\\
u_{e}  &  =u_{e0}+\varepsilon u_{e1}+\varepsilon^{2}u_{e2}+\varepsilon
^{3}u_{e3}+...,n_{h}=1+\varepsilon n_{h1}+\varepsilon^{2}n_{h2}+\varepsilon
^{3}n_{h3}+....\text{    .}\nonumber
\end{align}

Substituting (7) and (8) into the basic set of Eqs. (1)-(6), the first
order in $\varepsilon$ gives the following relations:%
\begin{equation}
n_{i1}=\frac{1}{V_{p}^{2}-3\sigma_{i}}\phi_{1},\text{ }%
u_{i1}=\frac{V_{p}}{V_{p}^{2}-3\sigma_{i}}\phi_{1},\text{
}n_{e1}=\frac{-\mu_{e}}{\left(  u_{e0}-V_{p}\right)  ^{2}-3\mu_{e}\sigma_{e}}\phi_{1},
\tag{9}%
\end{equation}%
\begin{equation}
u_{e1}=\frac{\mu_{e}\left(u_{e0}-V_{p}\right)}{\left(  u_{e0}-V_{p}\right)  ^{2}-3\mu_{e}\sigma_{e}}%
\phi_{1},\text{ }n_{h1}=\frac{-1+2k_{c}}{-3+2k_{c}}\phi_{1}. \tag{10}%
\end{equation}
and Poisson equation gives the following compatibility condition%
\begin{equation}
\frac{\mu_{e}\delta_{e}}{\left(  u_{e0}-V_{p}\right)  ^{2}-3\mu_{e}\sigma_{e}}+\frac
{\delta_{i}}{V_{p}^{2}-3\sigma_{i}}+a_{1}=0, \tag{11}%
\end{equation}
where $a_{1}=\left(  2/\left(  3-2k_{h}\right)  \right)  -1$. 
For the next-order in $\varepsilon$, we obtain a set of equations,
after making use of Eqs. (9)--(11), we obtain the final evolution equation in the form of KdV equation as
\begin{equation}
\frac{\partial\phi}{\partial\tau}+P\phi\frac{\partial\phi}{\partial
\xi}+Q\frac{\partial^{3}\phi}{\partial\xi^{3}}=0, \tag{12}%
\end{equation}
where%
\[
Q=\frac{2V_{p}\delta_{i}}{\left(  V_{p}^{2}-3\sigma_{i}\right)  ^{2}}%
+\frac{2(V_{p}-u_{e0})\mu_{e}\delta_{e}}{\left(  \left(  V_{p}-u_{e0}\right)
^{2}-3\mu_{e}\sigma_{e}\right)  ^{2}}.
\]%
\[
P=Q\left(  p_{1}+p_{2}\right)  ,\text{ \ }p_{1}=\frac{\left(  1+2k_{h}\right)
\left(  1-2k_{h}\right)  }{\left(  3-2k_{h}\right)  ^{2}},
\]%
\[
p_{2}=\frac{3\delta_{i}\left(  V_{p}^{2}+\sigma_{i}\right)  }{\left(
V_{p}^{2}-3\sigma_{i}\right)  ^{3}}-\frac{3\delta_{e}\mu_{e}^{2}(\left(
V_{p}^{2}-u_{e0}\right)  +\mu_{e}\sigma_{e})}{\left(  \left(  V_{p}%
-u_{e0}\right)  ^{2}-3\mu_{e}\sigma_{e}\right)  ^{3}}.
\]
and $\phi\equiv\phi_{1}$

\section{Solutions of the KdV equation}

Here, we are interested in discussing the possible solutions of the nonlinear
partial differential KdV equation, Eq. (12) that may fit the wave observations
by different space missions like the multi-spacecraft of Cluster, MMS, and Parker Solar Probe. The following parameters are adopted in
the whole figures where the density, temperature, and the velocity of the
electron beam are $n_{e}=0.003$ cm$^{-3}$, $T_{e}=300$ eV, $u_{e0}=0.55$ keV,
respectively. The superthermal electrons is characterized by a temperature of
$T_{h}=3000$ eV and density of $n_{h}=0.04$ cm$^{-3}$ \cite{r17}. We utilized
the direct integration method to find the solution of the KdV equation
depending on the sign of both the nonlinear coefficient and the dispersion
coefficient. Introducing the travelling wave solution, $\phi\left(  \xi
,\tau\right)  =\varphi\left(  \eta\right)  ,$ $\eta=\xi-M$ $\tau,$ where $M$
is the Mach number, in the KdV Eq. (12), then integrtated once, we obtain the
following ordinary differential equation:%
\begin{equation}
\frac{d^{2}\varphi}{d\eta^{2}}+P_{1}\text{ }\varphi^{2}-P_{2}\text{ }%
\varphi=0,\tag{13}%
\end{equation}
where
\[
P_{1}=\frac{P}{2Q},\text{ }P_{2}=\frac{M}{Q}.
\]
Integrating Eq. (13) and choosing the constant of integration to be zero, we obtain%
\begin{equation}
\left(  \frac{\varphi}{d\eta}\right)  ^{2}=P_{2}\text{ }\varphi^{2}%
-\frac{P_{1}}{6}\varphi^{3}.\tag{14}%
\end{equation}
The possible solutions of Eq. (14) depend on the polarity of the coefficients $P_{1}$ and $P_{2}$. Figure (1) presented the positive (red zone) and negative (yellow zone) values of the coefficient $P_{1}$. The coefficient $P_{2}$ had only positive values so we did not include the conture plot for it. 
Using the direct integration method for Eq. (14) considering
$P_{1}$ and $P_{2}$ are positive, we have the following solution form:%
\begin{equation}
\varphi\left(  \eta\right)  =A_{1}^{-1}\left(  \operatorname{sech}[\left(
\frac{P_{2}}{4}\right)  ^{1/2}\eta]\right)  ^{2},\tag{15}%
\end{equation}
where $A_{1}=P_{1}/\left(  6P_{2}\right)$. Equation (15) represents the
soliton wave. Similarly, for $P_{1}<0$ and $P_{2}>0,$, we have
\begin{equation}
\varphi\left(  \eta\right)  =A_{3}^{-1}\left(  \operatorname{csch}[\left(
\frac{P_{2}}{4}\right)  ^{1/2}\eta]\right)  ^{2},\tag{16}%
\end{equation}
where $A_{3}=P_{4}/\left(  6P_{2}\right)  ,$ $P_{4}\equiv-P_{1}$. This solution is known as explosive wave. The profiles of the soliton and explosive waves had depicted by Figs. (2a) and (2b), respectively. In the next section, we ignored the discussion for the explosive wave, since there is no space observation for it. 

\section{Structures of nonlinear ion acoustic solitary waves}

Now, it is convenient to discuss the impact of the variation of the
temperature ratio, the density ratio, and the normalized velocity on the
profile of the soliton wave. Figure (3) depicted the variation of the soliton
profile and the associated electric field with the normalized beam velocity.
It is clear that both of the positive and negative beam velocity give rise to the propagation of negative soliton waves according to Figs. (3a) and (3c), respectively. In the case of positive beam velocity, the width of the negative soliton pulse is wider than in the case of counter beam velocity. Moreover, increasing the positive normalized beam velocity (the case of
co-evolution) slightly alters the pulse amplitude, while increasing the negative
normalized beam velocity (the case of counter-evolution) leads to decrease the
pulse amplitude. 
The fluctuation of the soliton profile with the ratio of the ion-to-superthermal electron number density and the ratio of the ion-to-superthermal electron temperature is depicted in Figs. (6a) and (6c), respectively. Figure (6a) demonstrates that raising the number density ratio reduces the pulse amplitude slightly. As seen in Fig. (6c), increasing the temperature ratio causes the pulse width to narrow and the pulse amplitude to drop.
The influence of the normalised beam velocity, number density ratio, and temperature ratio on the maximum soliton amplitude is summarised in Fig. (7). In the case of positive co-evolution, as shown in Fig. (7a), the maximum soliton amplitude marginally increases, however it notably drops as the negative counter-evolution, $\delta_{i},$ and $\sigma_{i}$ increase, according to Figs. (7b), (7c) and (7d), respectively.
\section{Numerical results}
This section demonstrates the coincidence between the soliton solution and observed solitary waves in magnetospheric plasma and solar wind.

\subsection{ Earth's magnetosphere application}

The current work was propelled by the observation of electrostatic solitary
pulses at the separatrix zone at the magnetic reconnection in the near-Earth
magnetotail \cite{lim2014, r17}. The two stream instability is supposed to be
the mechanism behind the excitation of solitary waves. This is because the
process of magnetic reconnection causing the generation of an electron beam
beam which is able to excite the solitary pulses. Here are the cluster
observations for the background plasma and the electron beam $T_{h}=3000$ eV,
$n_{h}=0.04$ cm$^{-3}$, $T_{e}=300$ eV, $n_{e}=0.003$ cm$^{-3}$, and
$u_{e0}=0.4-1$ keV, where\ $T_{h}$ and $n_{h}$ are the background temperature
and density, while $T_{e},$ $n_{e},$ and $u_{e0}$ are the beam temperature,
density, and velocity, respectively. For example, utilizing these numerical
values in our model give rise to the formation of the peak-to-peak
unnormalized electric field, the potential is unnormalized by $k_{B}T_{h}/e$
$\approx3$ x $10^{6}$ mV and the space is unnormalized by $\lambda_{0}=\left(
k_{B}T_{h}/\varepsilon_{0}n_{h0}e^{2}\right)  ^{1/2}\approx2037$ m, with
amplitude $\approx$ 15 mV m$^{-1}$ for the counter electron beam as presented
by Fig. (8). This value is close to the observed values by the cluster on the
separatrix of the magnetic reconnection zone of the Earth's magnetotail \cite{r17}.
Similarly, the MMS observaions in the Earth's magnetopause reported the existence of electric field of $\sim 100 \mathrm{mV} \mathrm{m}^{-1}$ at the region of the magnetic reconnecion. Utilizing the plasma parameters $\left(T_{e} \sim 1     \mathrm{eV}\right.$, $T_{h} \sim 1 \mathrm{keV}, n_{e} \sim 0.2 \mathrm{~cm}^{-3}$, and $n_{h} \sim 30 \mathrm{~cm}^{-3}$ ) in our theoretical model yield an electric field of $E \approx 93 \mathrm{mV} \mathrm{m}^{-1}\left(k_{\mathrm{B}} T_{h} / e=10^{3} \mathrm{mV}\right.$ and $\lambda_{0}=43 \mathrm{~m}$ ), which is close to the observed value according to Fig. (7) \cite{mw}.

\subsection{Solar wind application}
Moreover, our model is suitable to investigate the observed solitary waves in
the solar wind. We modified the model to include the alpha particles instead
of the the electron beam. Considering the slow solar wind, the following
numerical values had been used: the ratio between the proton temperature and
the alpha particle temperature to the thermal electron temperature is less
that one. The ratio between the unperturbed number density of the alpha
particles to the unperturbed number density of the thermal electrons is has
the range of $\sim0.0-0.05$ and the ratio of the streaming alpha particles to
the electron acoustic speed has the range of $\sim0.0-0.3$ \cite{sing1, sing2,
sing3}. Figure (9) showed an example of the soliton pulse and the
corresponding electric field in the solar wind. Calculating the peak amplitude
of the electric field gives $\approx$ 0.25 mV m$^{-1}$ where the potential is
unnormalized by $\sim12$ x $10^{3}$ mV and the space is unnormalized by
$\sim115$ m. The measured electric field have the range of $\sim$
$0.0054-0.54$ mV m$^{-1},$ hence the current example is coincidence with the
solar wind observations.

\section{Conclusion}

We utilized the multifluid model to investigate the origin of solitary waves in the regions of magnetic reconnection in the Earth's magnetosphere and the solar wind. The KdV evolution equation has been derived by means of the reductive perturbation technique to examine the dynamics of ion acoustic solitary waves that are observed by different space missions such as cluster, MMS, and PSP. The core results can be summarized as:

\begin{itemize}
\item In the Earth's magnetotail, the current model predicts the formation of negative solitary that co-propagate with the electron beam positive velocity and counter
propagation with the electron beam negative velocity.

\item Increasing the positive electron beam velocity leads to a slight increase in the
soliton amplitude, while increasing the negative
electron beam velocity leads to decrease both the soliton amplitude and width.

\item Changing the ion to beam number density ratio and the ion to beam
temperature ratio significantly decrease the pulse amplitude.

\item In the solar wind, increasing the velocity of alpha particles leads to a
significant decrease in the soliton amplitude.
\end{itemize}

Finally, this work might be a possible diagnostic for the correlation between
the electron beam counter propagation with the stimulated soliton waves at the
separatrix of the magnetic reconnection at the Earth's magnetotail. Moreover,
we stressed that this model should be developed to address the origin of the
recent observations of electrostatic waves in the inner heliosphere
\cite{new1} and the role of whistler waves and electrostatic modes in super
accelerating alpha particles very close to the sun \cite{new2}.

\section*{Acknowledgments}
W. M. Moslem thanks the sponsorship provided by the Alexander von Humboldt
Foundation (Bonn, Germany) in the framework of the Research Group Linkage
Programme funded by the respective Federal Ministry.

\begin{figure}[h]
  \centering
  \begin{minipage}[b]{0.48\textwidth}
    \includegraphics[width=\textwidth]{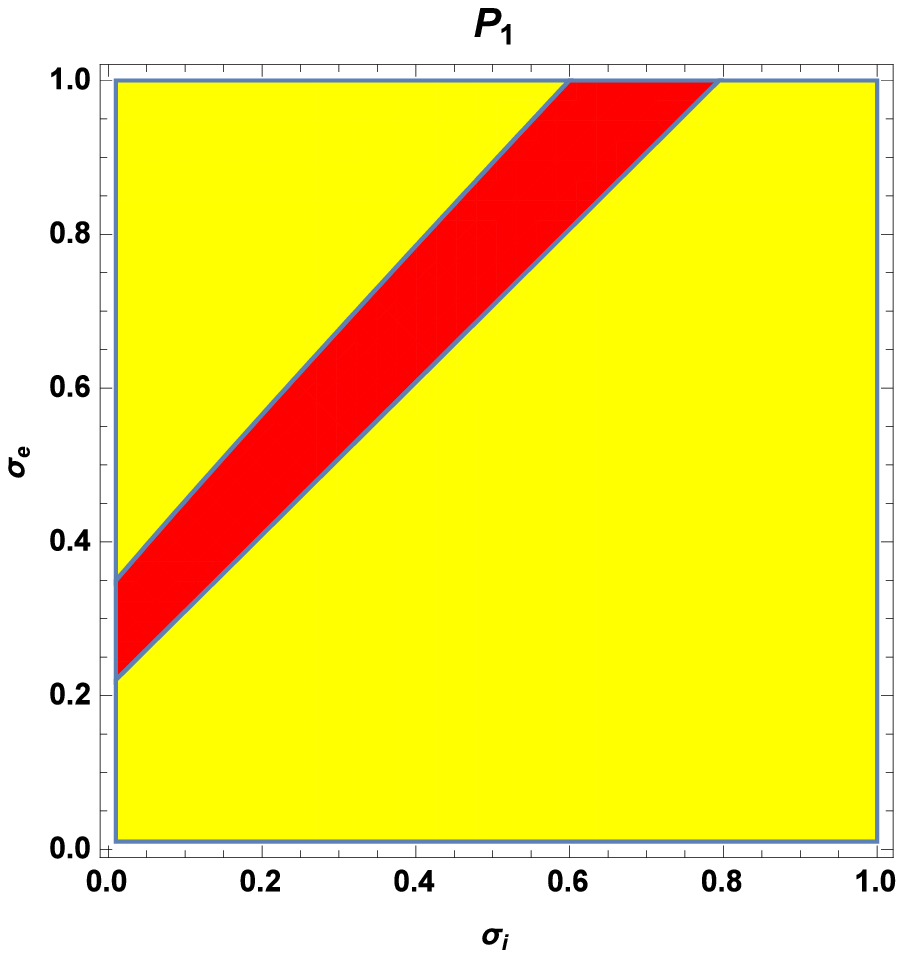}
  \end{minipage}
	\caption{ The possible values of $P_{1}$ is depicted against the
ion to superthermal electron temperature ratio $\sigma_{i}$ and the beam to superthermal electron temperature ratio $\sigma_{e}$ for $\delta_{e}=\delta_{i}-1,$
$M=0.3,k_{h}=3,$ $K_{B}=1.38\times10^{-23}$j k$^{-1},$ $m_{e}=9.109\times
10^{-31}$kg$,$ $n_{e0}=3\times10^{3}$m$^{-3},$ $n_{h0}=4\times10^{3}m^{-3},$
$u_{e0}=0.35,T_{e}=300$ eV$,$ $T_{i}=3000$ eV, $T_{h}=3000$ eV, $\mu
_{e}=1836.$}
\label{Figure7}
\end{figure}
\begin{figure}[h]
  \centering
  \begin{minipage}[b]{0.48\textwidth}
    \includegraphics[width=\textwidth]{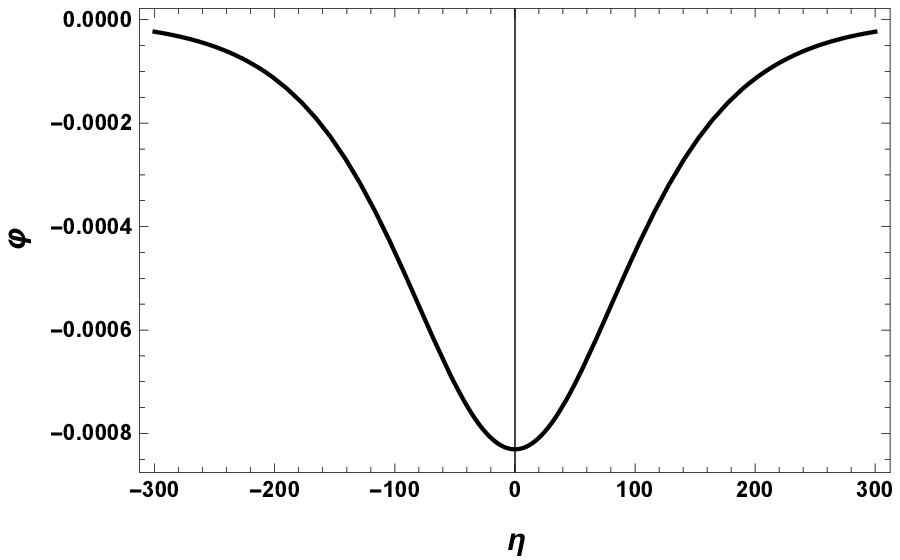}
  \end{minipage}
  \hfill
  \begin{minipage}[b]{0.48\textwidth}
    \includegraphics[width=\textwidth]{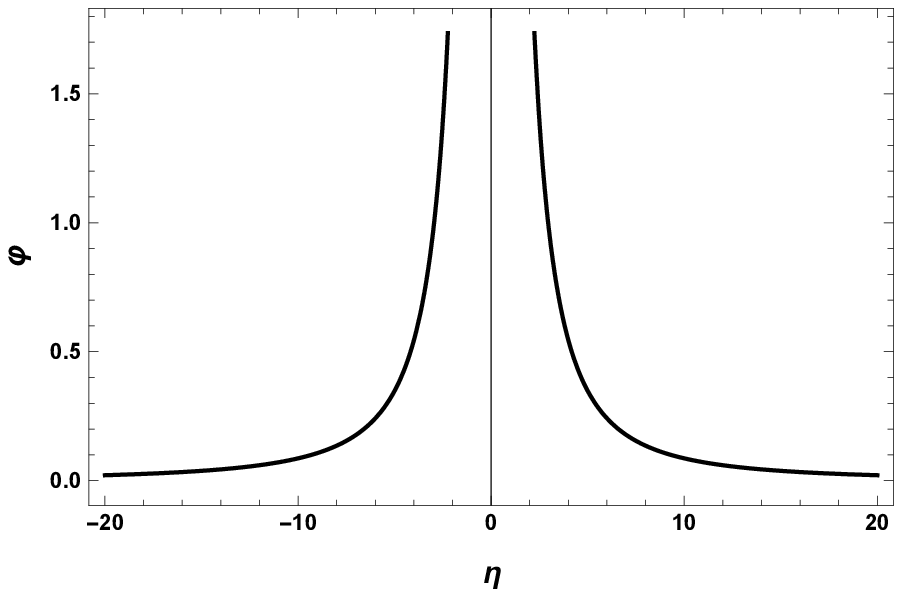}
  \end{minipage}
	\caption{(a): The two-dimensional profile of the ion
acoustic soliton wave of Eq. (15) is depicted against $\eta$. (b): The two-dimensional profile of the ion acoustic explosive wave of Eq. (16) is depicted against $\eta$. The same parameters were employed as in Fig (1).}
\label{Figure7}
\end{figure}
\begin{figure}[h]
  \centering
  \begin{minipage}[b]{0.45\textwidth}
    \includegraphics[width=\textwidth]{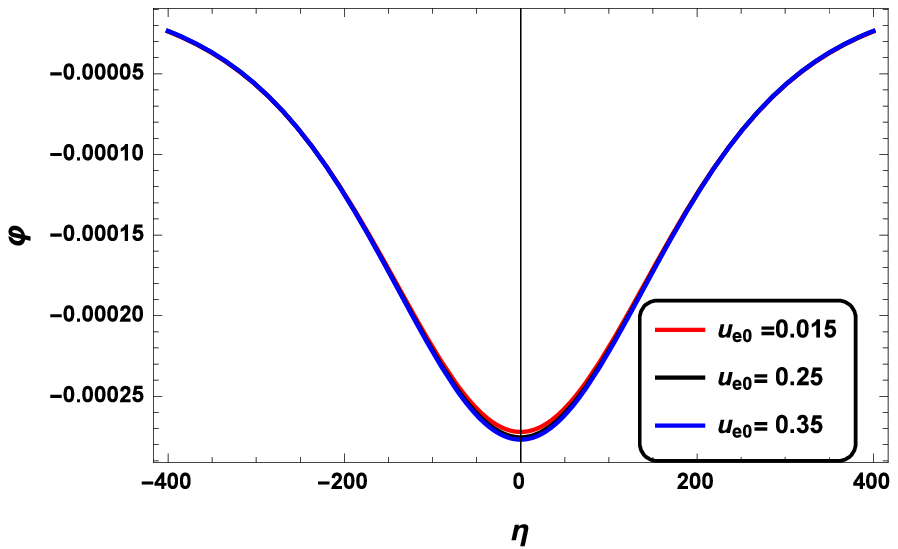}
  \end{minipage}
  \hfill
  \begin{minipage}[b]{0.45\textwidth}
    \includegraphics[width=\textwidth]{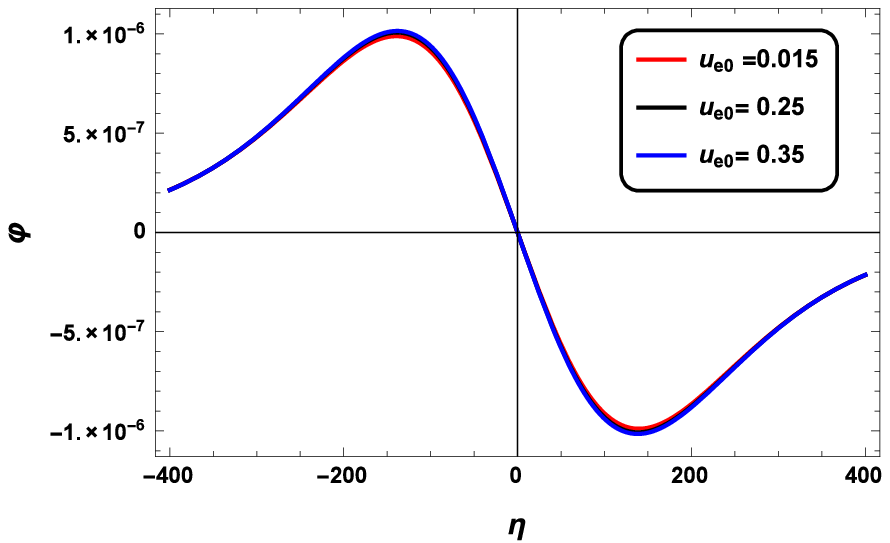}
  \end{minipage}
    \centering
	\begin{minipage}[b]{0.45\textwidth}
    \includegraphics[width=\textwidth]{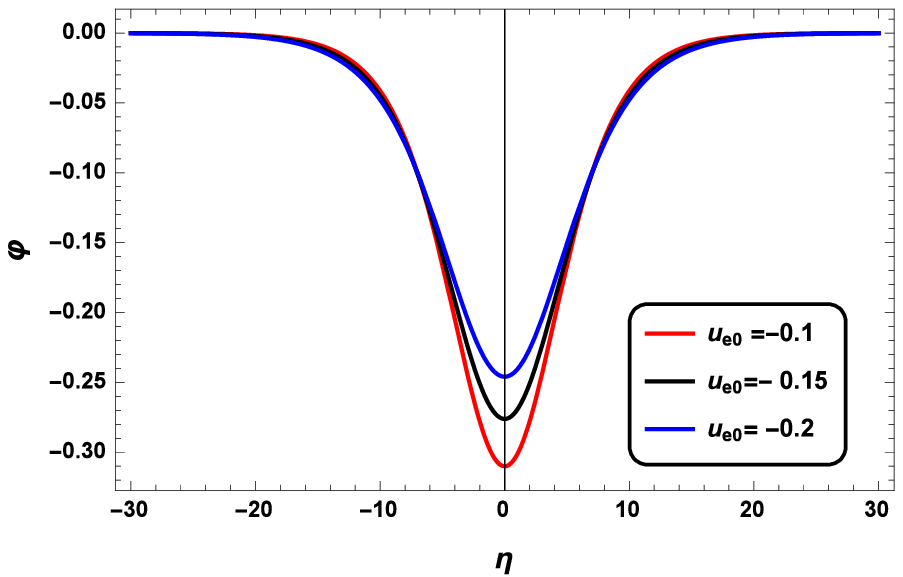}
  \end{minipage}
  \hfill
  \begin{minipage}[b]{0.45\textwidth}
    \includegraphics[width=\textwidth]{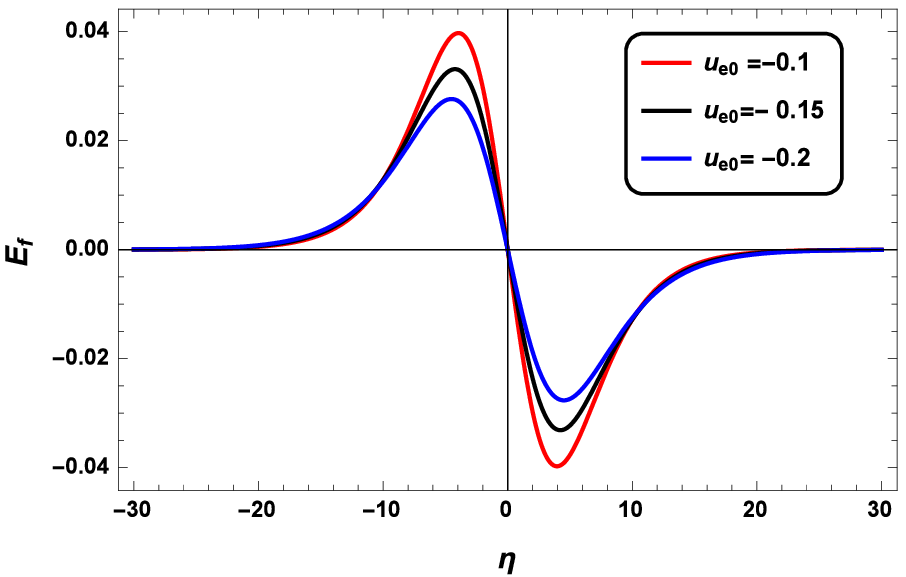}
  \end{minipage}
	\caption{Two-dimensional profile of
the\textbf{ }soliton ion acoustic wave of Eq. (15) depicted against $\eta.$
(a): For different values of positive $u_{e0}.$ (b): The associate electric
field of the positive soliton pulses (c): The variation of the negative
soliton pulse with the negative $u_{e0}.$ (d): The associate electric field of the negative soliton pulses$.$ Here, $M=0.1$ and $k_{h}=3$ for the positive solitons, while $M=0.3$ and $k_{h}=3$ for the negative solitons.}
\label{Figure9}
\end{figure}
\begin{figure}[h]
  \centering
  \begin{minipage}[b]{0.45\textwidth}
    \includegraphics[width=\textwidth]{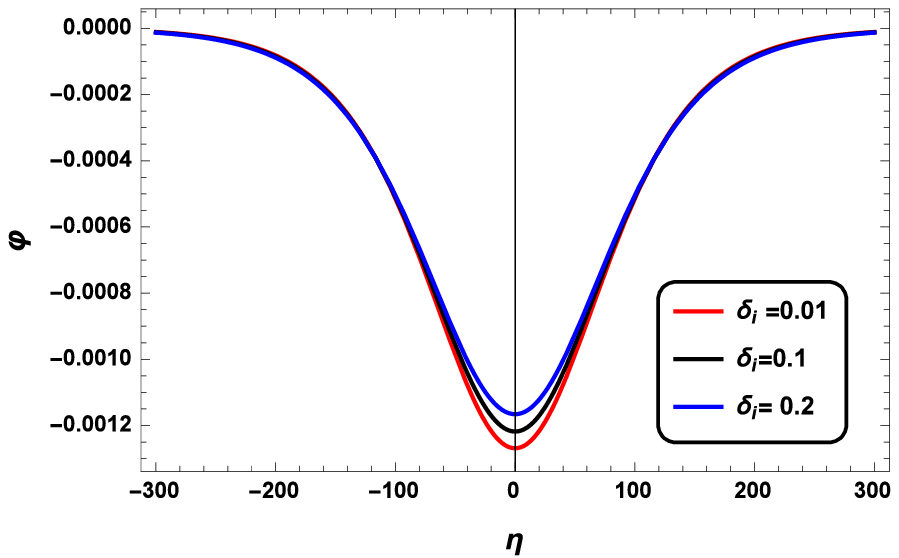}
  \end{minipage}
  \hfill
  \begin{minipage}[b]{0.45\textwidth}
    \includegraphics[width=\textwidth]{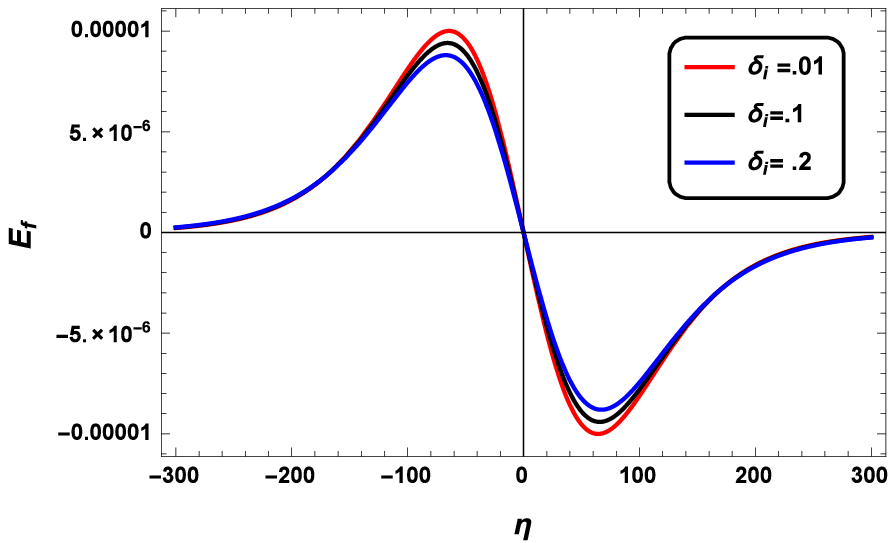}
  \end{minipage}
    \centering
	\begin{minipage}[b]{0.45\textwidth}
    \includegraphics[width=\textwidth]{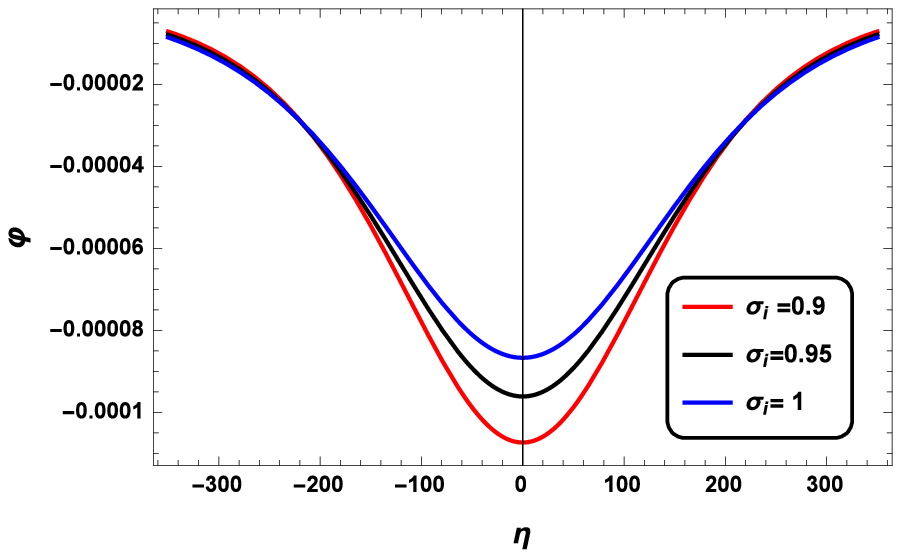}
  \end{minipage}
  \hfill
  \begin{minipage}[b]{0.45\textwidth}
    \includegraphics[width=\textwidth]{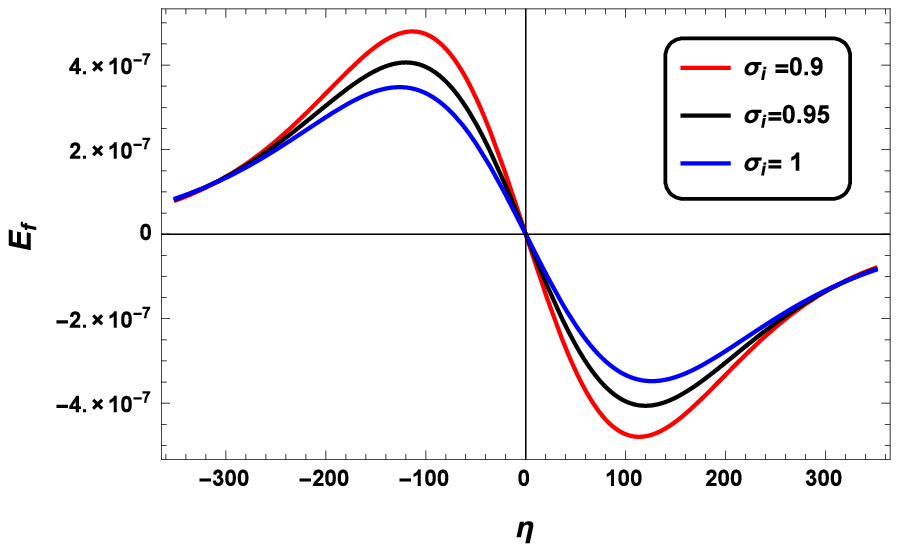}
  \end{minipage}
	\caption{The variation of the soliton profiles with the
plasma parameters: (a) the ion to beam number density ratio, and (b)
represents the associated electric field for $u_{e0}=0.35,$ $M=0.3$ and $k_{h}=3.$ While the impact of the ion to beam temperature ratio and the associated electric field had been depicted by (c) and (d), respectively for $u_{e0}=0.35,$ $M=0.1$ and $k_{h}=2.$} 
\label{Figure9}
\end{figure}
\begin{figure}[h]
  \centering
  \begin{minipage}[b]{0.45\textwidth}
    \includegraphics[width=\textwidth]{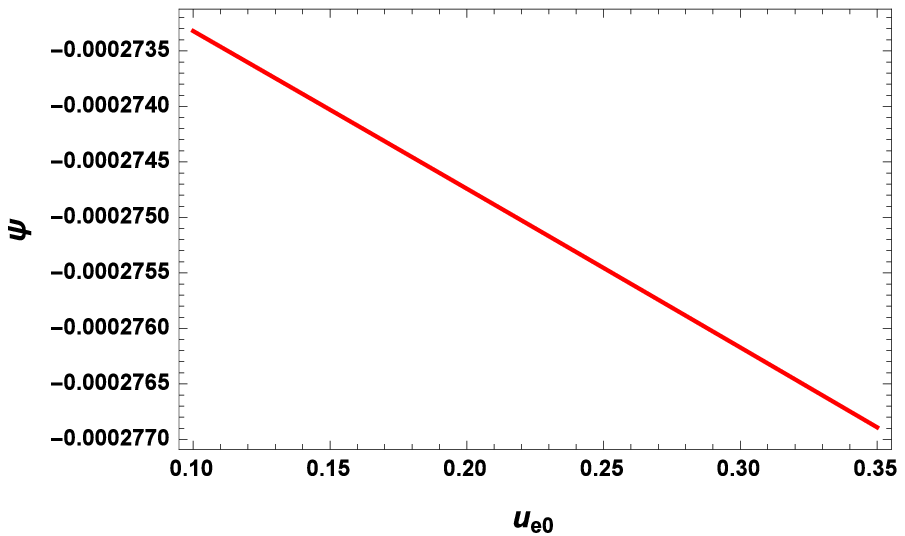}
  \end{minipage}
  \hfill
  \begin{minipage}[b]{0.45\textwidth}
    \includegraphics[width=\textwidth]{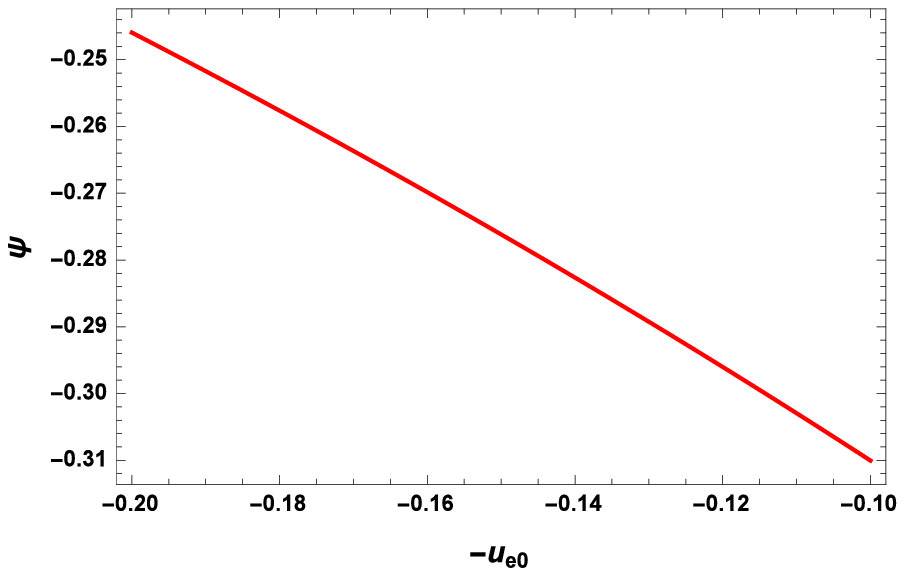}
  \end{minipage}
    \centering
	\begin{minipage}[b]{0.45\textwidth}
    \includegraphics[width=\textwidth]{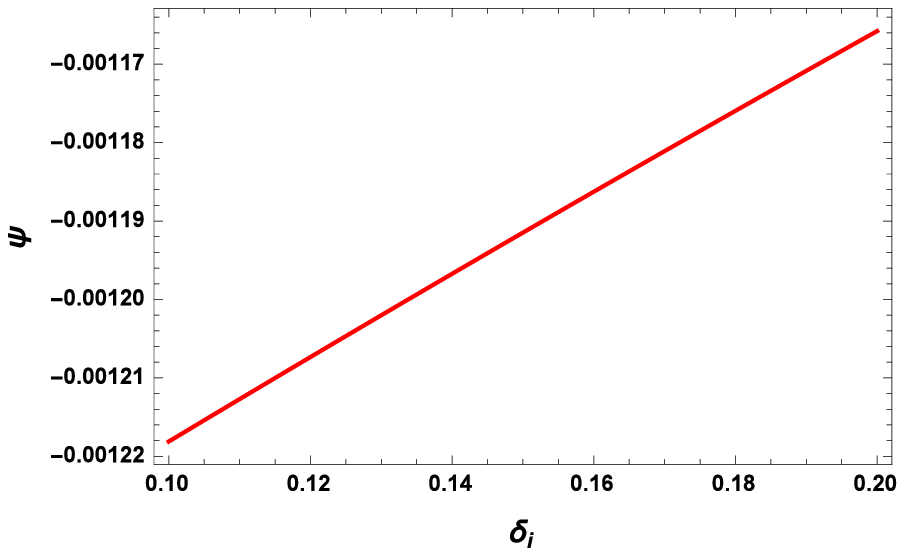}
  \end{minipage}
  \hfill
  \begin{minipage}[b]{0.45\textwidth}
    \includegraphics[width=\textwidth]{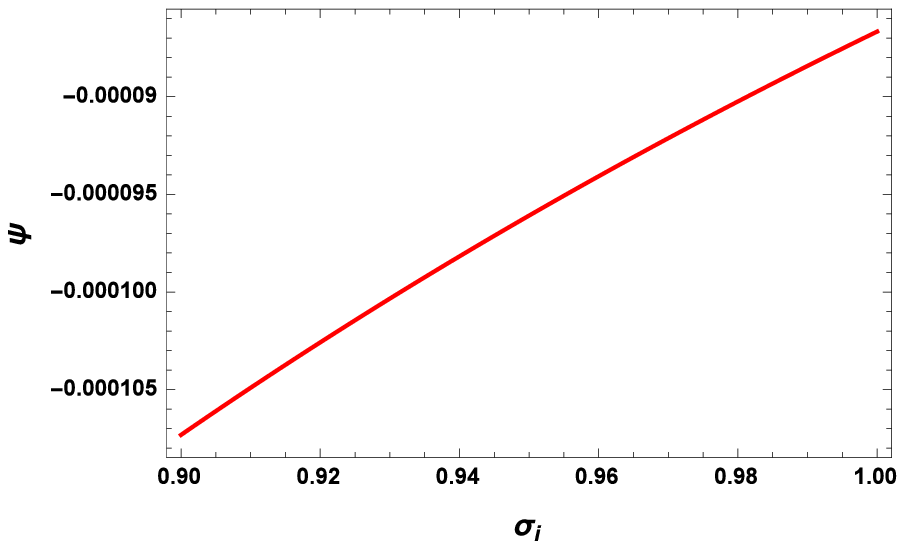}
  \end{minipage}
	\caption{The variation of the maximum soliton amplitude with (a) the
positive beam velocity where $M=0.1$ and $k_{h}=3.$, (b) the negative beam velocity where $M=0.3$ and $k_{h}=3.$, (c) the ion to beam number density ratio for $u_{e0}=0.35,$ $M=0.3$ and $k_{h}=3.$, and (d) the ion to beam temperature ratio for $u_{e0}=0.35,$ $M=0.1$ and $k_{h}=2.$.}
\label{Figure9}
\end{figure}
\begin{figure}[h]
  \centering
  \begin{minipage}[b]{0.48\textwidth}
    \includegraphics[width=\textwidth]{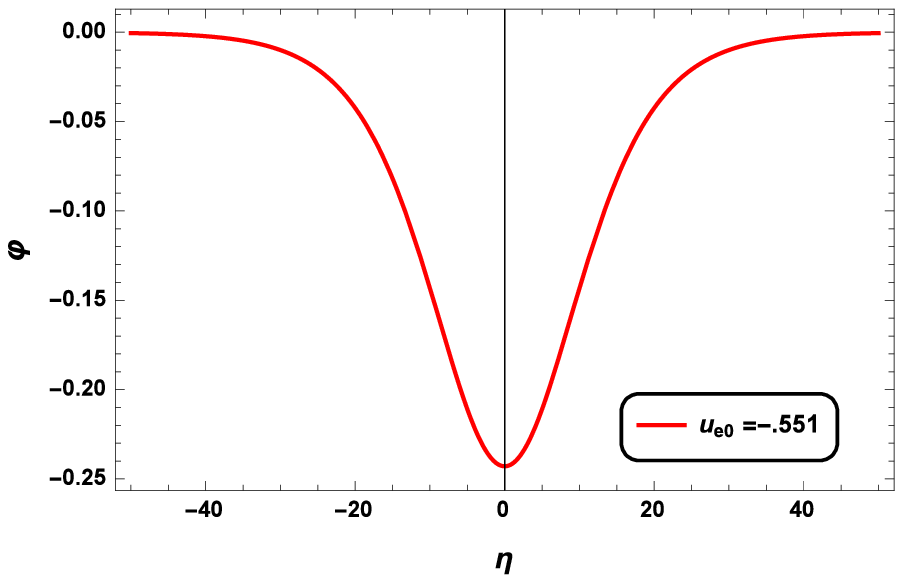}
  \end{minipage}
  \hfill
  \begin{minipage}[b]{0.48\textwidth}
    \includegraphics[width=\textwidth]{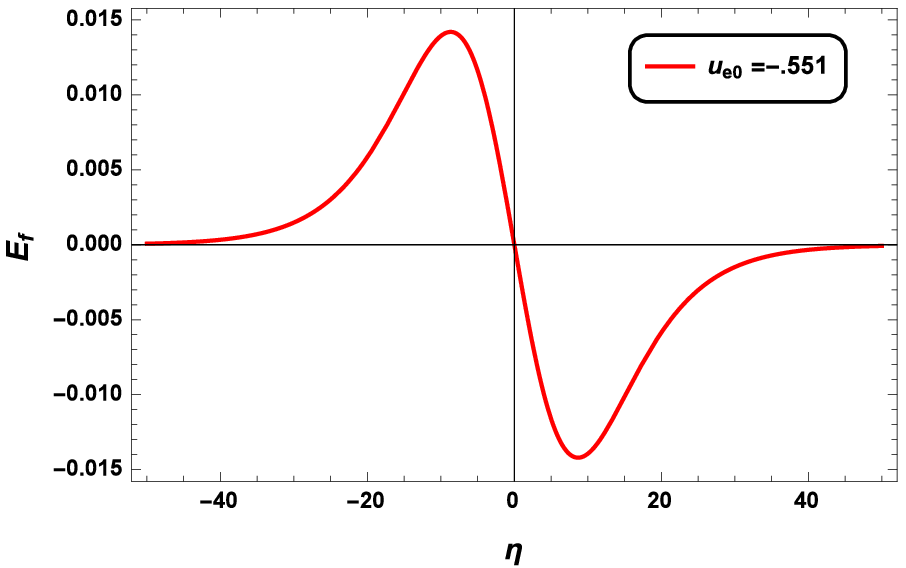}
  \end{minipage}
	\caption{(a) The soliton pulse agianst the negative electron beam
in the Earth's magnetotail. (b) For the associated electric field. Here,
$\delta_{i}=1.08,$ $M=0.03$ and $k_{h}=3.$}
\label{Figure7}
\end{figure}
\begin{figure}[h]
  \centering
  \begin{minipage}[b]{0.48\textwidth}
    \includegraphics[width=\textwidth]{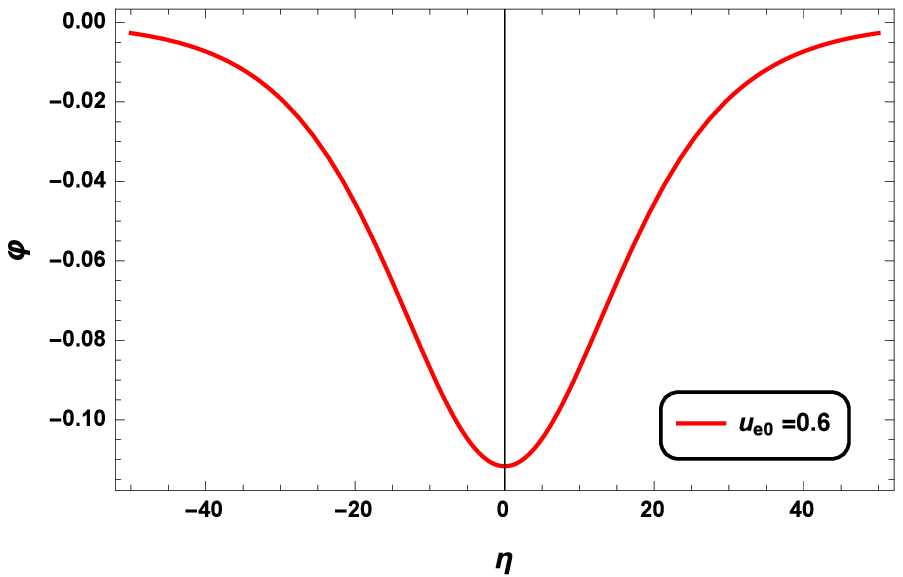}
  \end{minipage}
  \hfill
  \begin{minipage}[b]{0.48\textwidth}
    \includegraphics[width=\textwidth]{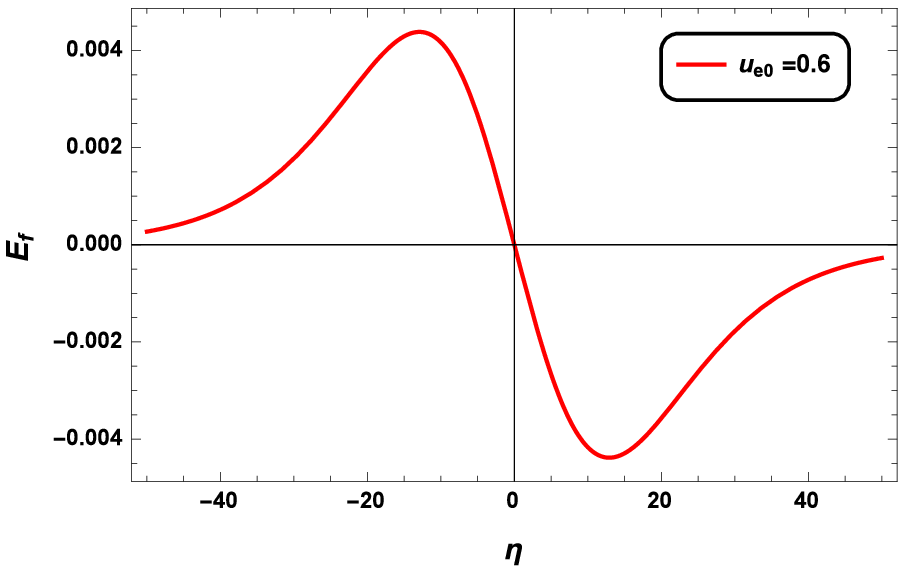}
  \end{minipage}
	\caption{(a) The soliton pulse agianst the electron beam velocity
in the Earth's magnetopause. (b) For the associated electric field. Here, $\delta_{i}=1.4,$ $M=4$ and $k_{h}=100.$}
\label{Figure7}
\end{figure}
\begin{figure}[h]
  \centering
  \begin{minipage}[b]{0.48\textwidth}
    \includegraphics[width=\textwidth]{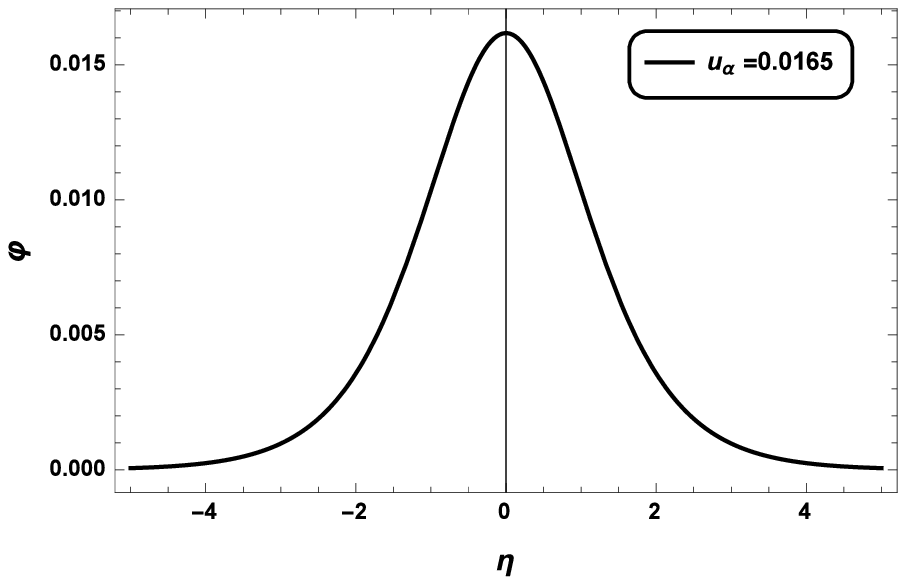}
  \end{minipage}
  \hfill
  \begin{minipage}[b]{0.48\textwidth}
    \includegraphics[width=\textwidth]{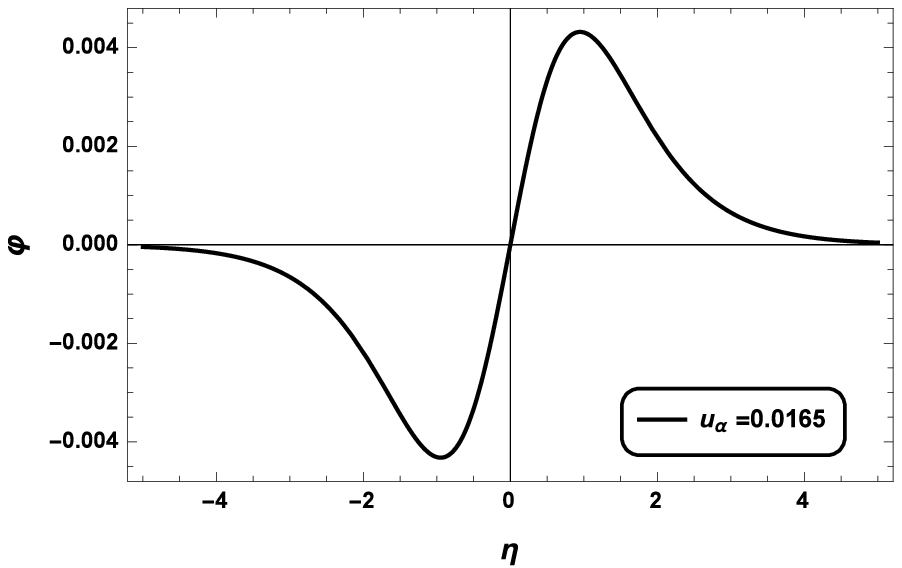}
  \end{minipage}
	\caption{(a) The soliton pulse agianst the electron beam velocity
in the solar wind. (b) For the associated electric field. Here, $\delta
_{i}=0.9,$ $M=0.0001$ and $k_{h}=3.$}
\label{Figure7}
\end{figure}
\end{document}